\documentclass{article}
\usepackage{spconf,amsmath,graphicx}
\usepackage{booktabs}
\usepackage{color}
\usepackage{pifont}
\usepackage[justification=centering]{caption}

\usepackage{enumitem}
\setlist{nosep, leftmargin=14pt}

\usepackage{mwe} 

\title{Adaptive PromptNet for auxiliary glioma diagnosis without contrast-enhanced MRI}

\name{Yeqi Wang$^{1, 2}$, Weijian Huang$^{2}$, Cheng Li$^{2}$, Xiawu Zheng$^{6}$, Yusong Lin$^{3, 4, 5 \star}$, Shanshan Wang$^{2, 6, 7 \star}$}

\address{
$^{1}$School of Computer and Artificial Intelligence, Zhengzhou University, Zhengzhou, Henan, China \\
$^{2}$Paul C. Lauterbur Research Center for Biomedical Imaging, Shenzhen Institutes of \\
Advanced Technology, Chinese Academy of Sciences, Shenzhen, Guangdong, China\\
$^{3}$School of Cyber Science and Engineering, Zhengzhou University, Zhengzhou, Henan, China\\
$^{4}$Collaborative Innovation Center for Internet Healthcare, Zhengzhou University, Zhengzhou, Henan, China \\
$^{5}$Hanwei IoT Institute, Zhengzhou University, Zhengzhou, Henan, China\\
$^{6}$Peng Cheng Laboratory, Shenzhen, Guangdong, China\\
$^{7}$Guangdong Provincial Key Laboratory of Artificial Intelligence in Medical Image \\ 
Analysis and Application, Guangdong Provincial People’s Hospital, \\
Guangdong Academy of Medical Sciences, Guangzhou, Guangdong, China
}

\begin{document}

\maketitle

\begin{abstract}
Multi-contrast magnetic resonance imaging (MRI)-based automatic auxiliary glioma diagnosis plays an important role in the clinic. Contrast-enhanced MRI sequences (e.g., contrast-enhanced T1-weighted imaging) were utilized in most of the existing relevant studies, in which remarkable diagnosis results have been reported. Nevertheless, acquiring contrast-enhanced MRI data is sometimes not feasible due to the patient's physiological limitations. Furthermore, it is more time-consuming and costly to collect contrast-enhanced MRI data in the clinic. In this paper, we propose an adaptive PromptNet to address these issues. Specifically, a PromptNet for glioma grading utilizing only non-enhanced MRI data has been constructed. PromptNet receives constraints from features of contrast-enhanced MR data during training through a designed prompt loss. To further boost the performance, an adaptive strategy is designed to dynamically weight the prompt loss in a sample-based manner. As a result, PromptNet is capable of dealing with more difficult samples. The effectiveness of our method is evaluated on a widely-used BraTS2020 dataset, and competitive glioma grading performance on NE-MRI data is achieved. 
\end{abstract}
\section{Introduction}
\label{sec:intro}

Gliomas are the most common type of primary neuroepithelial malignant tumors with high incidence and mortality rates, accounting for nearly 30\% of all primary brain tumors \cite{1ostrom2020cbtrus}. According to their histological appearance, gliomas can be divided into high-grade gliomas (HGG, WHO grade III and IV) and low-grade gliomas (LGG, WHO grade I and II). In clinical practices, the treatment and prognosis of LGG patients and HGG patients differ substantially \cite{3louis20162016, 4obara2020adult}. Hence, accurate gliomas diagnosis is critical for the development of correct therapeutic planning.

Multi-contrast magnetic resonance imaging (MRI), including non-enhanced MRI and contrast-enhance MRI, is a powerful imaging method that contributes significantly to auxiliary glioma diagnosis in the clinic \cite{5caulo2014data}. Visually inspecting the acquired MR images and distinguishing the grades of gliomas are time-consuming, which can only be performed by experienced physicians. Furthermore, with the widening employment of MRI, it becomes increasingly infeasible to manually analyze the rapidly accumulated volume of MRI data. Therefore, there is an urgent need to develop automatic MRI-based auxiliary diagnosis algorithms to assist physicians in performing accurate and timely glioma grading preoperatively.

\begin{figure*}[htb]
    \centering
    \includegraphics[width=0.9\textwidth]{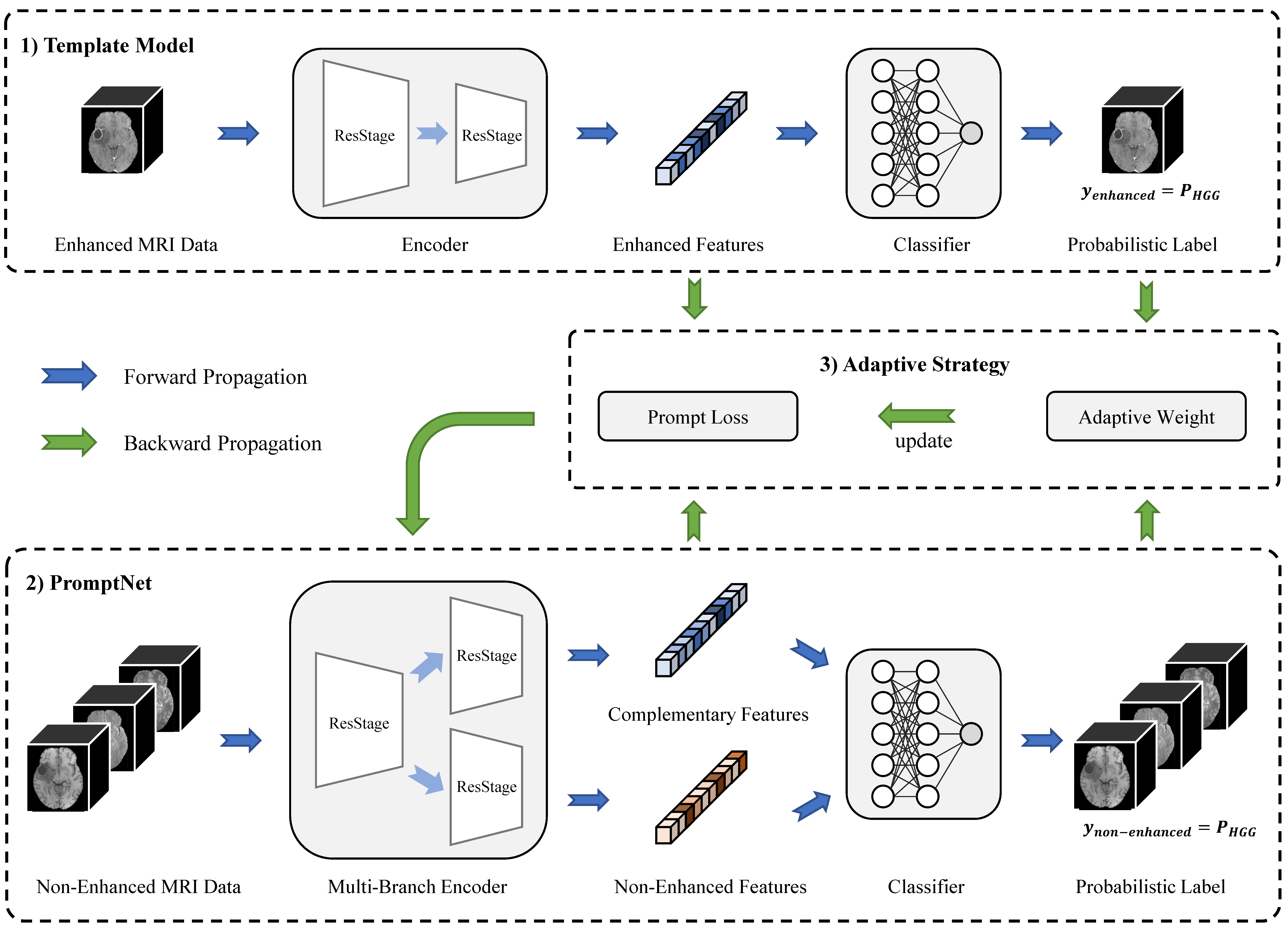}
    \caption{Schematic illustration of the proposed adaptive PromptNet.}
    \label{fig:my_label}
\end{figure*}


In the past decade, deep learning, especially convolutional neural networks (CNNs), has revolutionized the medical imaging field \cite{wang2016accelerating, wang2021annotation}. Existing studies have validated the capability of deep learning to conduct accurate glioma auxiliary diagnosis when contrast-enhanced MRI (CE-MRI) data are provided \cite{6ayadi2021deep, 7chelghoum2020transfer, 8deepak2019brain}. Nevertheless, it is more difficult to acquire CE-MRI data than NE-MRI data. Some patients may feel uncomfortable or even be allergic to the contrast agents. There are reports indicating that the most frequently utilized contrast agents of MRI, gadolinium-based contrast agents, may be linked to the occurrence of diseases, such as nephrogenic systemic fibrosis \cite{9khawaja2015revisiting}. Besides, the process of acquiring CE-MRI data is relatively longer, and the cost is also higher. To this end, some studies have developed algorithms to achieve MRI synthesis \cite{10huang2021brain} and lesion segmentation \cite{11zhou2021latent} without CE-MRI data. However, to the best of our knowledge, MRI-based automatic auxiliary glioma diagnosis while CE-MRI data are missing is rarely reported.

In this paper, we proposed an adaptive PromptNet for glioma grading, which requires only NE-MRI data for testing. The training of PromptNet is accomplished in two stages. In the first stage, a template model is trained on enhanced MRI data to extract enhanced features. In the second stage, our proposed PromptNet on non-enhanced MRI (NE-MRI) data is optimized with supervisory signals from both the glioma grading and the template model. PromptNet employs a multi-branch encoder to extract both non-enhanced features and complementary features. Through minimizing a designed prompt loss between the enhanced and complementary features, prompts from the template model can be transferred to PromptNet to help extract more useful information in the complementary feature extraction branch. To facilitate the optimization of PromptNet, a specific strategy is designed to adaptively update the weights of the prompt loss for each sample in each backward iteration. To summarize, the contributions of this work are three-fold:

\begin{enumerate}
    \item To avoid the reliance of current deep learning models on the CE-MRI data for glioma grading, a PromptNet is proposed along with a special designed loss. The model is designed to explore the correlations between different MR sequences and thus extracts inherent features of gliomas for accurate classifications even with only NE-MRI data during testing. 
    \item To balance the contributions from different samples, an adaptive weight updating strategy is designed for the prompt loss of each sample which can pay more attention to difficult samples during training.
    \item The effectiveness of the proposed PromptNet is evaluated on a public dataset, BraTS2020, and PromptNet generates better glioma grading results than the three state-of-the-art methods when testing on NE-MRI data.
\end{enumerate}

\section{Method}
\subsection{Problem Statement}
In this paper, we aim to construct a deep learning model for automatic glioma grading on NE-MRI data. Let $(x,y)$ be a pair of sample of NE-MRI data, where ${x}\in{X}$ refers to the image and $y\in{Y}$ is the corresponding glioma grade label. $Y$ represents the input data space, and $Y=[0,1]^k $ is the label space ($k$ is the number of classes). The automatic diagnosis model $f:X\rightarrow{R}_+$ maps $x$ to the probabilistic label $y'$. As shown in Fig. 1, the model $f$ can be divided into two phases. In the first phase, the encoder $f_E:X\rightarrow{R}^d$ maps the input $x$ to the latent feature vector $z:=f_E (x)$. In the second phase, the classifier, which consists of several full connected layers $f_{FC}:R^d\rightarrow{R}_+$, maps $z$ to the probabilistic label $y'$. A loss function $l:f(X)\times{Y}\rightarrow{R}_+$ is proposed to measure the glioma grading error according to the provided label $y$. The objective function can be defined as:
\begin{equation}
    f^*(x)={argmin}\ l(f(x),y)+\lambda{R(f)}
\end{equation}
where $R$ presents the regularization item, and $\lambda$ is a constant to balance the contributions of the two parts. 

\subsection{PromptNet}
PromptNet is proposed to address the difficulties of CE-MRI data acquisition. Particularly, PromptNet is trained with the help of CE-MRI data, but during testing, it requires only NE-MRI data. 
The optimization of PromptNet involves two stages (Fig. 1). In the first stage, a template model is trained using CE-MRI data to extract contract-enhanced features $z_{CEF}$. In the second stage, these enhanced feature are employed as prompts to help guide the feature extraction of PromptNet. To prevent the overwhelming effects of enhanced features, a multi-branch encoder to extract non-enhanced features and extra complementary features $z_{CF}$ is constructed for PromptNet. The enhanced features influence only the complementary feature extraction procedure explicitly through a prompt loss $l_{prompt}$:
\begin{equation}
    l_{prompt}=|z_{CEF}-z_{CF}|
\end{equation}

Then, non-enhanced features and complementary features are fused and inputted to a classifier to obtain the auxiliary glioma diagnosis results. The total objective function for the training of PromptNet can be defined as:
\begin{equation}
    f^*(x)=argmin\ l(f(x),y)+l_{prompt}
\end{equation}

\subsection{Adaptive Prompt Loss Weight Updating Strategy}
Since MRI data of different patients are heterogeneous, the potential of each sample to serve as the prompt is different. To this end, a sample-level performance gap-based adaptive prompt loss weight updating strategy is designed for PromptNet. Here, we define the template model as $f_{T}$, which maps the CE-MRI data $x_{CE}$ to the corresponding probabilistic label. The CE-MRI data and NE-MRI data from the same glioma patient sample are fed to the template model and PromptNet, respectively. The PromptNet adjusts the weight of the prompt loss dynamically during training according to the discrepancy between the obtained probabilistic labels. In this way, more attention can be paid to the difficult samples. We adopt the $L1$ norm to calculate the performance gap:
\begin{equation}
    w_{adaptive} = |f_{T}(x_{CE})-f(x)|.
\end{equation}
The objective function of the adaptive PromptNet is defined as follows:
\begin{equation}
    f^*(x)={argmin}\ l(f(x),y)+w_{adaptive}\times l_{prompt}.
\end{equation}

\section{Experiment}
\subsection{Dataset}
The Brain Tumor Segmentation Challenge (BraTS) 2020 dataset \cite{12menze2014multimodal} is utilized in this study to evaluate the automatic auxiliary glioma diagnosis performance of different methods.
The dataset contains 369 glioma patient samples with two glioma grades (LGG and HGG). 
We randomly divided these samples into a training dataset and a test dataset with a ratio of 4:1. 
MRI data from four sequences are provided for each sample, including T1, T1CE, T2, and FLAIR. 
All the brain MRI data have been properly pre-processed by the challenge organizers \cite{12menze2014multimodal}.

\subsection{Implementation Details}
The structure of the proposed adaptive PromptNet is shown in Fig. 1. Two major components are involved, including a multi-branch encoder and a classifier. The former includes a stem cell and three 3D residual stages, extracting latent feature vectors from 3D brain MRI data. In addition, we adopt group normalization and GELU activation to adapt to the small batch size. The classifier consists of three full connected layers.

We train all of our models using Adam optimizer for 100 epochs. The initial learning rate is set to1e-5, and it decays by a factor of 10 at the 30th and 60th epochs. The batch size is 4. All models are trained on a NVIDIA TITAN V GPU with 12GB memory. All work related to deep learning is based on the open-source platform TensorFlow v2.7.0. Each experiment is repeated five times. The area under the receiver operating characteristic curve (AUC) is utilized as the main evaluation metric. Other metrics, including accuracy, precision, sensitivity, and the area under the precision-recall curve (PRC), are also calculated.

\subsection{Results}

\begin{table*}[!htbp]
    \centering
    \renewcommand\arraystretch{1.5}
    \begin{tabular}{ccccccc}
        \toprule [2pt]
            \textbf{Type} & \textbf{Data} & \textbf{AUC} & \textbf{PRC} & \textbf{Accuracy} & \textbf{Precision} & \textbf{Sensitivity} \\
        \midrule
            3D CNN & FS-MRI & 0.9706±0.0141 & 0.9926±0.0036 & 0.9250±0.0153 & 0.9526±0.0206 & 0.9458±0.0076 \\
        \midrule
            2D CNN \cite{6ayadi2021deep} & NE-MRI & 0.8330±0.0376 & 0.9550±0.0103 & 0.7922±0.0362 & 0.8622±0.0172 & 0.8802±0.0340 \\
            TL \cite{7chelghoum2020transfer} & NE-MRI & 0.8457±0.0198 & 0.9560±0.0064 & 0.8208±0.0116 & 0.8363±0.0119 & 0.9651±0.0108 \\
            TL+ML \cite{8deepak2019brain} & NE-MRI & 0.8576±0.0019 & 0.9505±0.0005 & 0.8216±0.0060 & 0.8358±0.0054 & \textcolor{red}{\textbf{0.9661}}±0.0001 \\
           \textbf{Our Method} & NE-MRI & \textcolor{red}{\textbf{0.9451}}±0.0080 & \textcolor{red}{\textbf{0.9872}}±0.0019 & \textcolor{red}{\textbf{0.9000}}±0.0118 & \textcolor{red}{\textbf{0.9448}}±0.0137 & 0.9254±0.0093\\
        \bottomrule[2pt]
        \end{tabular}
    \caption{Auxiliary glioma diagnosis results of different methods. FS-MRI represents full sequence MRI data.}
    \label{tab:my_label1}
\end{table*}

\begin{table*}[!htbp]
    \centering
    \renewcommand\arraystretch{1.5}
    \begin{tabular}{ccccccc}
        \toprule [2pt]
            \textbf{$l_{prompt}$} & \textbf{$w_{adaptive}$} & \textbf{AUC} & \textbf{PRC} & \textbf{Accuracy} & \textbf{Precision} & \textbf{Sensitivity}  \\
        \midrule
            \ding{56} & \ding{56} & 0.9203±0.0166 & 0.9810±0.0040 & 0.8625±0.0153 & 0.9162±0.0275 & 0.9085±0.0390 \\
            \ding{52} & \ding{56} & 0.9379±0.0098 & 0.9850±0.0026 & 0.8763±0.0273 & 0.9118±0.0251 & \textcolor{red}{\textbf{0.9390}}±0.0093 \\
            \ding{52} & \ding{52} & \textcolor{red}{\textbf{0.9451}}±0.0080 & \textcolor{red}{\textbf{0.9872}}±0.0019 & \textcolor{red}{\textbf{0.9000}}±0.0118 & \textcolor{red}{\textbf{0.9448}}±0.0137 & 0.9254±0.0093\\
        \bottomrule[2pt]
        \end{tabular}
    \caption{Results of ablation studies of the proposed adoptive PromptNet.}
    \label{tab:my_label2}
\end{table*}

\begin{figure}[!htb]
    \centering
    \includegraphics[width=0.40\textwidth]{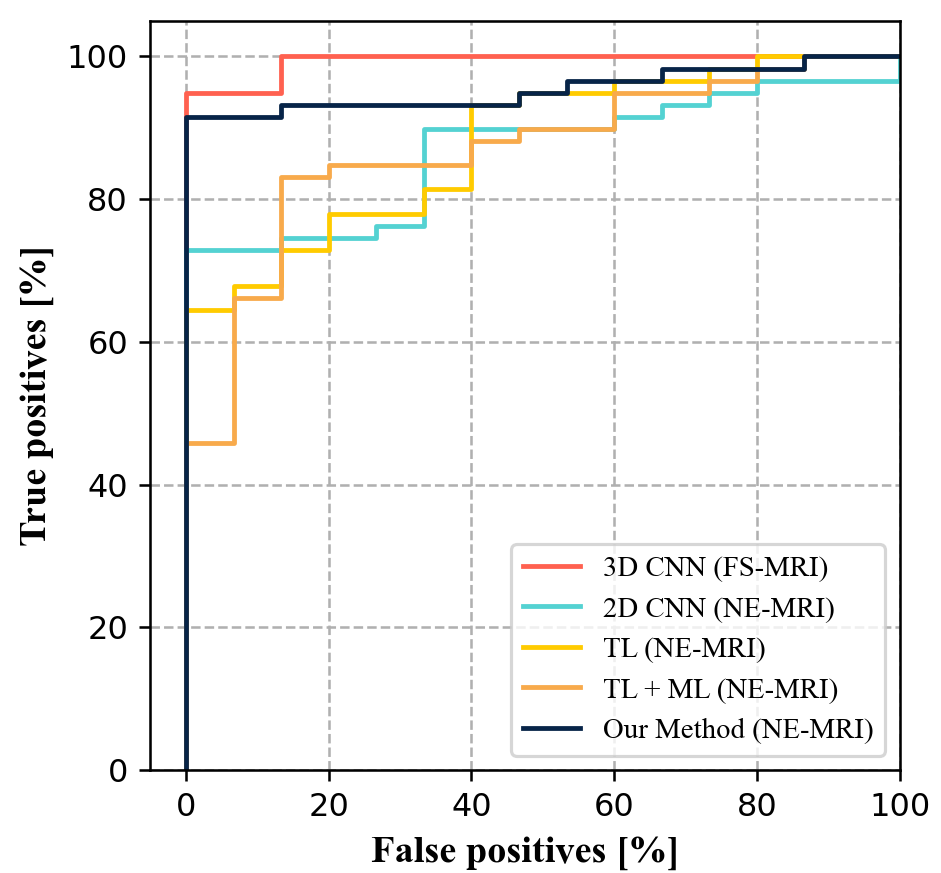}
    \caption{ Receiver operating characteristic curve of different methods}
    \label{fig:my_label}
\end{figure}

We evaluate the performance of the proposed adaptive PromptNet utilizing the three NE-MRI sequences (T1, T2, and FLAIR). 
Our method is compared to the following brain tumor classification algorithms: 
1) Transfer Learning (TL): An ImageNet-pretrained ResNet-50 is finetuned with brain MRI data. 
2) Transfer Learning + Machine Learning (TL+ML): An ImageNet-pretrained ResNet-50 is employed to extract deep features, and a SVM classifier is trained on these features to conduct the classification. 
3) 2D CNN: A custom CNN model is trained on brain MRI data from scratch. 
Specifically, we extract 20 2D slices from each 3D brain MRI data. 
4) 3D CNN: A 3D CNN model is trained on full sequence data (T1, T1CE, T2, and FLAIR), and the model has the same network structure as the template model in Fig. 1.

Table.1 lists the glioma grading results of the different methods. Our proposed PromptNet generates better auxiliary glioma diagnosis results than other existing methods when only NE-MRI data are provided. Although the 3D CNN tested on all four sequence MRI data achieves the best classification results, our method presents a high potential for glioma grading on NE-MRI data, which can be very important for real-world clinical applications. Fig. 2 plots the receiver operating characteristic curve of each method for direct comparisons. Similar observations can be made that our method is better than TL, TL+ML, and 2D CNN.

Table. 2 gives the results of the ablation studies related to the proposed method. These experiments were conducted to demonstrate the contributions of the prompt loss function and the adaptive prompt loss weight updating strategy. The results suggest that both the prompt loss and the adaptive strategy can improve the grading performance of the proposed PromptNet, and the best results are obtained by adopting both mechanisms. 

\section{Conclusion}
In this paper, we propose an adaptive PromptNet for auxiliary glioma diagnosis utilizing NE-MRI data. A prompt loss is designed and minimized to help capture more representative MRI information. Moreover, a performance gap-based adaptive strategy is proposed to adjust the contribution of the prompt loss to the network parameter optimization in a sample-based manner, and thus, the model can focus more on the difficult samples. Our proposed adaptive PromptNet has a high potential to be employed in real-world applications when CE-MRI data are infeasible to collect.

\section{Acknowledgement}
This research was partly supported by Scientific and Technical Innovation 2030 - "New Generation Artificial Intelligence" Project (2020AA\ A0104100, 2020AAA0104105), the National Natural Science Foundation of China (61871371, 62222118, U22A2040, 81772009), Guangdong Provincial Key Laboratory of Artificial Intelligence in Medical Image Analysis and Application (2022B1212010011), the Basic Research Program of Shenzhen (JCYJ20180507182400762), Shenzhen Science and Technology Program (RCYX20210706\ 092104034), Youth Innovation Promotion Association Program of Chinese Academy of Sciences (2019351), Collaborative Innovation Major Project of Zhengzhou (20XTZX06013, 20XTZX05015).

\bibliographystyle{IEEEbib}
\bibliography{strings,refs}

\end{document}